\documentclass[aps,preprint,superscriptaddress]{revtex4}
\usepackage{bm}
\usepackage{epsfig,amsopn}
\usepackage{graphicx}
\usepackage{footnote}
\usepackage{amsmath,amssymb}
\usepackage{natbib}
\usepackage{verbatim}
\renewcommand{\section}[1]{{\par\it #1.---}}
\def\be{\begin{equation}}
\def\ee{\end{equation}}
\def\bea{\begin{eqnarray}}
\def\eea{\end{eqnarray}}
\def\la{\langle}
\def\ra{\rangle}
\def\om{\omega}
\def\nn{\nonumber}
\def\p{\partial} 
\def\mJ{{\mathcal{J}}}
\def\f{\frac}

\def\hM{\hat{M}}
\def\hT{\hat{T}}
\def\hG{\hat{\Gamma}}
\def\hu{\hat{u}}
\def\hv{\hat{v}}
\def\hz{\hat{z}}
\def\hd{\hat{d}}

\begin{document}
\title{ Heat conduction in  disordered harmonic lattices with energy conserving
  noise} 
\author{Abhishek Dhar}
\email{dabhi@rri.res.in}
\affiliation{Raman Research Institute, Bangalore 560080, India}
\author{K. Venkateshan}
\author{J.L. Lebowitz}
\affiliation{Departments of Mathematics and Physics, 
    Rutgers University, Piscataway, NJ 08854}
\date{\today} 
\begin{abstract}
We study heat conduction in a harmonic crystal whose bulk dynamics is
supplemented by random reversals (flips) of the velocity of each
particle at a rate $\lambda$. The system is maintained in a
nonequilibrium stationary state(NESS) by contacts with white noise Langevin
reservoirs at different temperatures.  
We show that the one-body and pair correlations in this system are the
same (after an appropriate mapping of parameters)  as those obtained
for a model with self-consistent reservoirs. This is true both for the
case of equal and random(quenched)  masses. While the heat
conductivity in the NESS of the ordered system is known explicitly,
much less is known about the random mass case. Here we investigate the
random system, with velocity flips. We improve the bounds on the
Green-Kubo conductivity obtained by Bernardin \cite{bernardin09}  . The conductivity of
the 1D system is then studied both numerically and analytically. This
sheds some light on the effect of noise on the transport properties of
systems with localized states caused by quenched disorder. 
 \end{abstract}
\maketitle

\section{Introduction}

We consider   heat transport in mass disordered harmonic lattices with
stochastic bulk dynamics.  
For the $1D$ disordered harmonic lattice without stochasticity the effect of localization due to disorder
leads, in 
the presence of pinning,  to an exponential decay of the heat
current as a function of the length \cite{dhar01,dharleb08}.  In the
absence of pinning the conductivity depends on the boundary conditions
either growing as $\sqrt{N} $ or decaying as $1/ \sqrt{N}$
\cite{casher71, rubin71}. The situation is very different when one
adds stochasticity to the 
dynamics. Bernardin \cite{bernardin09} obtained finite positive upper and lower bounds  on the
Green-Kubo conductivity of a  harmonic lattice with periodic boundary conditions subjected to  
stochastic dynamics which conserves energy but not momentum.

%Recently a model was introduced where the Hamiltonian dynamics of the harmonic
%lattice was augmented by local energy and momentum conserving
%stochastic dynamical rules \cite{basile06}. It was shown exactly that in the
%absence 
%of a pinning potential the Green-Kubo
%conductivity $\kappa_{GK}$  diverged . It was also shown
%that a harmonic pinning potential was sufficient to give a finite
%$\kappa_{GK}$. A numerical study of the nonequilibrium case of the
%above model, with slightly diferent local rules than those studied in
%\cite{basile06}, was done in \cite{delfini08} and they arrived at similar
%conclusions for the thermal conductivity $\kappa$ (defined through
%Fourier's law). 

In this paper we study the heat flux of such a
disordered harmonic system,   both pinned and unpinned, connected to white noise
Langevin type heat reservoirs with different temperatures at the two
ends. The stochastic part of the bulk dynamics consists of random reversals of 
each particle's velocity at a rate $\lambda$.
The analytical as well as accurate numerical tractability of
this model makes it a good test system to address the
problem of the effect of noisy dynamics on transport properties of
disordered systems. To the extent that stochastic dynamics affect
phonon-phonon interactions in some 
rough sense similar to anharmonicity, this may also teach us something
about the effects of the latter. 
The effect of interactions on localization
and transport in disordered systems has generated much interest
and has been  studied both for  the case of electron and
phonon transport
\cite{basko08,vadim07,pikovsky07,kopidakis07,dharleb08,dhar08,
bernardin09,mulansky09,vadim09,basko10}~.   
A study by two of us \cite{dharleb08} of the heat conductivity of the
disordered  1D pinned lattice with anharmonicity
found that a small amount of anharmonicity was sufficient to cause  a
transition to a diffusive regime with a finite value of $\kappa$. The
question of whether the transition from an insulator to conductor
occurs at zero or some finite small value of anharmonicity remains an
open problem. For the noisy dynamics on the other
hand \cite{bernardin09} shows that an arbitrarily small noise  leads to normal
transport.  The work of \cite{basko10} suggests a transition at zero
nonlinearity for classical systems (the quantum situation may be different). 

There have been  some recent studies \cite{delfini08,lepri09} on the NESS of ordered harmonic chains
with noisy dynamics conserving both energy and momentum, and with white noise
Langevin baths at different temperatures at the two ends. 
It was shown in these studies that the time-evolution of the
pair-correlations formed closed sets of equations. This closure is
true also for the energy conserving model that we study. In addition
we show a mapping of the steady state equations for one-body and
pair-correlation functions to that of the self-consistent reservoirs
model first introduced 
by  Bolsterli {\emph{etal}} \cite{BRV70,RV75} and recently solved exactly for
the ordered case by Bonetto {\emph{etal}} \cite{BLL04}.  

The plan of the paper is as follows. In sec \ref{model-mapping} we define the
model precisely and show the mapping to the model with self-consistent
reservoirs. 
In sec \ref{green-kubo}
we use the method of Bernardin \cite{bernardin09} to obtain improved lower and
upper bounds for the Green-Kubo thermal conductivity of the random mass system. In
sec \ref{numerics} we present results from numerical calculations as
well as nonequilibrium simulations for the dependence of the heat flux in the random mass case on
system size and on noise
strength. Both the pinned and unpinned system are studied. 
For the pinned case conductivity decreases with the strength of the binding
and increases with strength of the noise. For the unpinned case we study the
effect of boundary conditions (BCs)    \footnote{We find that the smaller the
  strength of the noise the larger $N$ has to be to have agreement between the
  different BCs. This leaves open the question of what happens if we first let
  $ N \rightarrow \infty$ and then make the noise strength go to zero.}.
Finally we conclude with a discussion of the nature of the NESS for this model. While it is easily checked that the NESS is not strictly Gaussian, we find that the one particle momentum and position distributions are very close to Gaussian distribution. Furthermore their values at different sites or of momentum and position at the same site are essentially uncorrelated.

\section{Noisy dynamics and self-consistent reservoirs} 
\label{model-mapping}
The results in this section apply in any dimension. For simplicity of notation we consider
here explicitly the one dimensional case; a harmonic chain with the Hamiltonian:
\bea
H &=& \sum_{l=1,N} \left[ \f{p_l^2}{2 m_l} + k_o \f{q_l^2}{2}\right]  
+  \sum_{l=2,N} k \f{(q_l-q_{l-1})^2}{2}  + k' \left[ \f{
  q_1^2}{2}+\f{q_N^2}{2}\right] \label{ham} \\
&=& \f{1}{2} \left[ p \hat{M}^{-1} p + q \hat{\Phi} q\right]~, \nn
\eea
where $\{q_l,p_l\}$ denote the position and momenta of the
particles. We have used the notation 
$p=(p_1,p_2,...,p_N),~q=(q_1,q_2,...,q_N)$ and $\hM$ and $\hat{\Phi}$ are
$N\times N$ matrices corresponding to masses  and forces respectively.
When $k_o > 0$ we have the pinned case and set $k'=k$. In the unpinned case, $k_o=0$, we
consider fixed, $k'>0$, and free, $k'=0$, boundary conditions.    

The system's evolution has a deterministic part described by the
Hamiltonian above and a stochastic part consisting of two different
processes: (i) every particle is subjected to a noise which flips
its momentum, {\emph{i.e.}} for the $l^{\rm th}$ particle the transition $p_l
\to -p_l$ occurs with a rate $\lambda$,  (ii) the particles at the
boundaries $l=1$ and $l=N$ are attached to heat baths with Langevin
dynamics at temperatures $T_L$
and $T_R$ respectively.
Thus the end particles have additional
terms in their equation of motion  of the form 
%AD: Changed following a bit
$-\gamma p_{\alpha}/m_\alpha+(2 \gamma T_{\alpha})^{1/2} \eta_\alpha(t)$, for $
\alpha =1,N$, with $\la \eta_\alpha(t) \eta_{\alpha'}(t') \ra =
\delta_{\alpha,\alpha'} ~\delta(t-t')$, $\gamma$ is 
the friction constant and $T_1=T_L,~T_N=T_R$ are the bath temperatures.

The master equation  describing the 
time evolution of the full phase space probability density 
is therefore given by:
\bea
\f{\p P(x)}{\p t} = \sum_{l,m} {\hat{a}}_{lm} x_m \f{\p}{\p x_l}  P +
\sum_{l,m} \f{\hat{d}_{lm}}{2} \f{\p^2 P}{\p x_l \p x_m} +\lambda \sum_l
    [P(...,-p_l,...)-P(...,p_l,...)]~,  \label{masteqn}
\eea
where $x=(q_1,q_2,\ldots,q_N,p_1,p_2,\ldots,p_N)=(x_1,x_2,\ldots,x_{2N}) $ and 
$\hat{a}$ and $\hat{d}$ are $2N \times 2N$ matrices given by:
\bea
\hat{a}=\left( \begin{array}{cc}
 0 & -\hM^{-1} \\
\hat{\Phi} & \hM^{-1} \hat{\Gamma}^{-1} \end{array} \right)
~~~~~\hat{d}= \left( \begin{array}{cc}
0 & 0 \\
0 & 2 \hat{T} \hat{\Gamma} \end{array} \right)~.
\eea
Here $\hat{T}$ and $\hat{\Gamma}$ are diagonal matrices with diagonal 
elements given by $\hat{T}_{ll}=T_L \delta_{l,1}+T_R \delta_{l,N}$ and
$\hat{\Gamma}_{ll} = \gamma (\delta_{l,1}+ \delta_{l,N})$
respectively. Similar to the case studied in \cite{delfini08} we also
find that the equations for the one-body and pair correlation
functions of the system are closed.  (In fact there are closed equations for each order of the correlation.)
We define the vector $\rho$ , $ \rho_{l}=<x_{l}>, \, l=1,2 \dots 2N$
and the pair correlation matrix  
\bea
\hat{c}=\left( \begin{array}{cc} 
\hu & \hz \\    \label{matc}
\hz^T & \hv \end{array} \right).  
\eea  
%AD: Changed: Please always put symbols such as N and numerical numbers
%within $$. Also please leave spaces after full stops!
where the $N \times N$ matrices $\hat{u} , \hat{z}$ and $\hat{v}$ are given by $\hat{u}_{lm}=\la q_l q_m \ra
~,\hat{v}_{lm}=\la p_lp_m \ra $ and $\hat{z}_{lm}=\la q_l p_m \ra$ .
It follows then from Eq.~(\ref{masteqn}) $\rho$ and $\hat{c}$ satisfy
the following equations of motion:  
\bea
\f{d \rho}{dt}&=& -\hat{a}\rho + \left( \f{d \rho}{dt}\right)_{col}~, \nn \\ 
\f{d\hat{c}}{dt} &=& -\hat{a} \hat{c} - \hat{c} \hat{a}^T +\hd + \left(
\f{d\hat{c}}{dt}\right)_{col}~, 
\eea 
where the last terms in the above two equations arise from the flip
dynamics and are given by:
\bea
\left( \f{d \rho}{dt} \right)_{col} &=& - 2 \lambda \left( \begin{array}{c}
  0 \\ \la p \ra \end{array} \right) \nn \\ 
\left( \f{d \hat{c}}{dt}\right)_{col} &=& - 2 \lambda \left( \begin{array}{cc} 0 & \hz
  \\ \hz^T & 2 ( \hv-\hv_D ) \end{array} \right)~, \label{corr}
\eea
and $\hv_D$ is a diagonal matrix with matrix elements $[\hat{v}_D]_{ll}=
\hat{v}_{ll}=\la p_l^2 \ra$. 

In the steady state, $d\rho/dt=0$  which implies
$\rho=0$.  Setting $d \hat{c}/dt=0$ gives the following set of equations
for the pair correlations in the NESS:  
\bea
&&\hz^T=-\hM \hz \hM^{-1}~, \nn \\
&&\hv=\f{1}{2}(\hM \hu \hat{\Phi} + \hat{\Phi} \hu \hM) +\f{1}{2} (
\hM \hz \hG \hM^{-1} +\hM^{-1} \hG \hz^T \hM )~, \nn \\
&&(\hM \hu \hat{\Phi} - \hat{\Phi} \hu \hM) + (\hM \hz \hG \hM^{-1} -
\hM^{-1} 
\hG \hz^T \hM ) + 2
\lambda ( \hM \hz - \hz^T \hM) = 0~, \nn \\
&&(\Phi \hz+ \hz^T \hat{\Phi}) + (\hM^{-1} \hG \hv + \hv \hG \hM^{-1})
+ 4 \lambda (
\hv-\hv_D) = 2 \hT \hG~. \label{nesseqs}
\eea
Using the fact that $\hu$ and $\hv$ are symmetric matrices we have
$N^2+N(N+1)$ unknown variables and there are that many independent
equations above.

Now consider the case of heat conduction across a  harmonic chain
with Hamiltonian given by ~(\ref{ham}) and self consistent reservoirs attached to all sites. This is in addition to
the two end reservoirs at fixed temperatures $T_L$ and $T_R$.
Each of the side reservoirs is a Langevin bath with a
friction constant $\gamma'_{l}$ and  a temperature $T'_l$, $l=1,2,...N$, which is self-consistently fixed by the condition that there is no net flow of
energy into the reservoir \cite{BLL04}.
The stochastic  equations of motion of this system are:
%AD: Corrected missing power of 1/2
\bea
\f{d p_1}{dt}&=&-\Phi_{1m}q_m-\f{\gamma}{m_1}p_1-\f{\gamma'_1}{m_1}p_1+(2
\gamma T_L)^{1/2} \eta_1(t) + (2 \gamma'_1 T'_1)^{1/2} \zeta_1(t) \nn \\ 
\f{d p_{l}}{dt}&=&-\Phi_{lm}q_m-\f{\gamma'_l}{m_l}p_l+ (2 \gamma'_l T'_l)^{1/2}
\zeta_l(t) ~~l=2,...,N-1~,\nn   \\
\f{d p_{N}}{dt}&=&-\Phi_{Nm}q_m-\f{\gamma}{m_N}p_N-\f{\gamma'_N}{m_N}p_N+(2
\gamma T_R)^{1/2} \eta_N(t) + (2 \gamma'_N T'_N)^{1/2} \zeta_N(t)~,     \label{selfconst}
\eea
where $\eta_1,\eta_N$ and $\zeta_l,~l=1,2,...,N$ are independent Gaussian
white noise sources with unit variance. It is immediately established that the
probability distribution $P(x)$ in the NESS of this model is a Gaussian. 
The self consistency condition for zero current into the side
%AD: Corrected missing subscript ``l'' below.
reservoirs is given by $T_l'=\hat{v}_{ll}=\la p_l^2 \ra/m_l$. Making the
identification $\gamma_l'=2 \lambda m_l$, it is  
seen that the equations for the pair-correlations in the steady state
corresponding to the above equations are given precisely by
Eq.~(\ref{nesseqs}). \\ 
The self-consistent model was first studied by  Bolsterli {\emph{et al}} 
\cite{BRV70} who introduced the self-consistent
reservoirs as a simple scattering mechanism mimicking anharmonicity and which
might ensure local equilibration and  the validity of Fourier's law.   
The model was  later solved exactly by Bonetto et al \cite{BLL04}
who proved approach to local equilibrium and validity of Fourier's law
for the ordered case, i.e where all the $m_{l} $'s are equal. They also 
obtained an explicit expression for the thermal conductivity of the system in
all dimensions. (see Eq. (\ref{kGK}) below)

\section{Bounds on Green-Kubo conductivity }
\label{green-kubo}
Bernardin \cite{bernardin09}  considered a  model of a disordered
harmonic chain with a stochastic noise that changes the momentum of
neighboring particles while keeping the sum of their kinetic energies
constant. He obtained an exact result for the Green-Kubo conductivity of an
ordered chain and also rigorous upper and lower bounds for the conductivity of
disordered chains. Here we use Bernardin's approach for our model to obtain an
exact expression for the ordered chain. We also 
obtain bounds for the conductivity of the disordered chain which are slightly
improved from those of Bernardin's. 

The time evolution of the phase space density is given by Eq. (\ref{masteqn}) which we rewrite here in a more abstract form for convenience.
\bea
\f{\p P(x)}{\p t} &=& L P(x) \nn \\
{\rm where}~~L &=& A + \lambda S \nn \\
AP(x) &=& \sum_{l=1}^{N} [- \f{p_l}{m_l} \f{\p P(x)}{\p q_l}  + \sum_{m=1}^{N} \Phi_{lm} q_{m}\f{\p P(x)} {\p p_l}]
\nn \\ 
SP(x) &=& \sum_l  [P(...,-p_l,...)-P(...,p_l,...)]~.  
\eea
The total current which is carried entirely by the Hamiltonian part can be written in the following form:
\bea
\mJ =\f{k}{2} \sum_{l=1}^{N} \f{p_l}{m_l} (q_{l+1}-q_{l-1}) \; \; ,  q_{0}=q_{N} \; , q_{N+1}=q_{1}. \label{totJ}
\eea

The Green-Kubo expression for the thermal conductivity at temperature T is given by:
\bea
\kappa_{GK}&=&\lim_{z \to 0} \lim_{N \to \infty} \f{1}{N T^2 } \int_0^\infty
  dt~e^{-z t}~ \la \mJ(0) \mJ(t) \ra  \nn \\
&=& \lim_{z \to 0} \lim_{N \to \infty} \f{1}{N T^2 } \int_0^\infty dt~e^{-zt}~ \int dx~ \mJ~ e^{L t}~( ~\mJ~ P_{eq}~ ) \nn \\
 %&=&\lim_{z \to 0} \lim_{N \to \infty} \f{1}{N T^2 } \int dx ~\mJ
  %~(z-L)^{-1}~(~ \mJ~ P_{eq}~ ) \nn
     &=& \lim_{z \to 0} \lim_{N \to \infty} \f{1}{N T^2 }  ~\la ~\mJ, ~(z-L)^{-1}~\mJ~ \ra~.  \label{GK}
\eea
\\
where we have used  the notation $\la f,g \ra= \int dx f(x) g(x) P_{eq}$ for
any two functions $f,g$ of phase space variables  $x=(q_{1},
\ldots,q_{N},p_{1},\ldots, p_{N})$ and $P_{eq} \sim exp[-\beta H]$ where $H$
is given by the periodicized version of Eq.~(\ref{ham}) with  $k'$
set equal to $0$.

We note the following relations which are  easy to prove:
\bea
A \mJ P_{eq}&=& \sum_{l,j} \f{\Phi_{lj}q_j}{m_l} (q_{l+1}-q_{l-1}) P_{eq} \nn \\
{\rm and}~~~ S\mJ P_{eq} &=& -2 \mJ P_{eq}~ \label{opax} 
\eea

\subsection{Green-Kubo conductivity for equal mass ordered case}
\label{subsec:eqmass}
For the equal mass case Eq.~(\ref{opax}) gives $A \mJ P_{eq}=0$. This
is true with or without pinning and corresponds to the fact that for periodic boundary conditions the 
current operator commutes with the Hamiltonian.  Hence we get:
\bea
\kappa_{GK} &=& \lim_{z \to 0} \lim_{N \to \infty} \f{1}{T^2 N} \int dx ~\mJ~ \f{1}{z+2 \lambda}~ \mJ ~P_{eq}
= \lim_{N \to \infty} \f{\la \mJ^2 \ra}{2 \lambda T^2 N} ~. \nn
\eea
Using the form of $\mJ$ in Eq.~(\ref{totJ}) we then get:
\bea
\kappa_{GK} &=& \f{kD}{8 \lambda m}~, \label{kGK} \\  
{\rm where}~~D &=& \f{4k}{2k+k_o+[(k_o)(4k+k_o)]^{1/2}}~. \nn
\eea
This agrees with the result of \cite{BLL04} [Eq.~(4.18)] where the
conductivity was defined as $\kappa = \lim_{N\to\infty} \la J_N \ra
N/(T_L-T_R)$ where 
%AD: Corrected formula below
$\la J_N \ra = {k}\la q_{l}p_{l+1}/m_{l+1} \ra$ ($l=1,2,\ldots,N-1$)  is the average heat flux  in the NESS of the system with
self-consistent reservoirs specified by (\ref{selfconst}). 
The mapping between the noisy dynamics model and the self-consistent
reservoirs model (with the transformation $\gamma=2 \lambda$m ) implies that
for our noisy model  also $\kappa=\kappa_{GK}$ in the ordered case.
Following the methods in \cite{BLL04} it is easy to show that the value of
$\kappa$ is independent of boundary conditions for the ordered case and while
not proven we expect this to be true also for the disordered case for $N
\rightarrow \infty$ at fixed $\lambda >0$.  In fact there is every reason to
believe that whenever the Green-Kubo formula for $\kappa_{GK}$ converges to a
finite value when $ N \rightarrow \infty$ then it will agree with the
conductivity in the NESS defined as $\kappa = \lim_{N \rightarrow \infty}
\lim_{T_{L} \rightarrow T_{R} \rightarrow T}  N \la J_{N} \ra /( T_L -T_R)$.

\subsection{Upper and lower bounds on the Green-Kubo conductivity} 
We now consider the random case where the masses are independently chosen from
some distribution. Bernardin's proof that the conductivity $\kappa_{GK}$ is bounded away from zero and infinity is based on an identity between $\la \mJ, (z-L)^{-1} \mJ \ra$ and a variational expression (Eq (15) in \cite{bernardin09}), 
\bea
\la \mJ, (z-L)^{-1} \mJ \ra = Sup\{ 2 \la u, J \ra - \la (z- \lambda
S) u, u \ra - \la (z-\lambda S)^{-1} A u, Au \ra \} \label{gkbound}
\eea
where the supremum is carried out over the set of smooth functions
$u(q,p)$.
The derivation of this formula is straightforward for $A$=0.  More generally  we can consider a symmetric $L$, e.g one corresponding to a stochastic dynamics satisfying detailed balance with respect to $P_{eq}$. Then we have that $u$ is the solution of the equation 
$Lu= \mJ$ , and both sides of  Eq.(\ref{gkbound}) are equal to $ \la \mJ, u \ra$.  For a derivation of Eq. (\ref{gkbound}) in the case $L=S+A$ with $A$ antisymmetric, see \cite{CLTSunder}. \\
\textbf{Lower bound}: Choose a test function $u=\mu \sum_l p_l
(q_{l+1}-q_{l-1})$, where $\mu$ is a variational parameter:
\bea
\la u, \mJ \ra &=& \f{k \mu  T }{2} \sum_l \la (q_{l+1}-q_{l-1})^2 \ra = N
T^2 \f{   \mu D}{2}~, \nn \\
\la (z-\lambda S) u,u \ra &=& (z+2 \lambda) \la u^2 \ra 
= (z+2 \lambda) \mu^2 T \sum_l m_l \la (q_{l+1}-q_{l-1})^2 \ra~. \nn
\eea
 where D is defined in Eq.~(\ref{kGK}).
 Denoting
by $[...]$ an average over disorder we then get:
\bea
\left[ \la (z-\lambda S) u,u \ra \right]  = N T^2 (z+2 \lambda) \f{\mu^2 D
  [m]}{k}~ \nn~.
\eea 
Similarly, 
\bea
\la (z-\lambda S)^{-1} A u, Au \ra = (z+4 \lambda)^{-1} \la (A u)^2
\ra 
=(z+4 \lambda)^{-1} \mu^2 T^2 \sum_l \left( \f{1}{m_l}-\f{1}{m_{l+1}}
\right)^2 m_l m_{l+1}~, \nn 
\eea  
and averaging over disorder gives
\bea
\left[ \la (z-\lambda S)^{-1} A u, Au \ra \right] = 2 N T^2 (z+4 \lambda)^{-1} \mu^2
\left( [m]\left[\f{1}{m}\right]-1\right)~.   \nn
\eea
Thus we have:
\bea
\f{1}{N T^2} [\la \mJ, (z-L)^{-1} \mJ \ra ] &\geq& D \mu -C \mu^2
\\ 
{\rm where~~~} 
C &=& \f{2 \lambda D  [m]}{k} +\f{1}{2 \lambda}([m]\left[\f{1}{m}\right]-1)~. 
\eea 
The minimum of the bound occurs at $\mu=D/(2 C)$ and this gives:
\bea
[ \kappa_{GK} ] \geq \f{D^2}{4 C}~.   \label{eqlb}
\eea

\textbf{Upper bound}: By neglecting the last term in Eq.~(\ref{gkbound}) which is clearly negative,  we
get the upper bound:
\bea 
\la \mJ, (z-L)^{-1} \mJ \ra &\leq &(z+2 \lambda)^{-1} \la \mJ^2 \ra =
(z+2 \lambda)^{-1}  T \f{k^2}{4} \sum_l \f{1}{m_l} \la
(q_{l+1}-q_{l-1})^2 \ra~.    \\
{ \rm Hence,}  \nn \\
~~[ \kappa_{GK}] &=& \f{1}{NT^2} \left[ \la \mJ, (z-L)^{-1} \mJ \ra
  \right]  \leq   \f{kD }{8 \lambda} [\f{1}{m}]~. \label{equb}
\eea
Combining (\ref{eqlb}) and (\ref{equb}) gives:
\bea
\f{kD}{8\lambda [m](1+ k\f{[1/m] -1/[m]}{4 \lambda^2 D})}  \leq  [\kappa_{GK}] \leq  \f{kD}{8 \lambda} [\f{1}{m}]
 \eea
As $\lambda \rightarrow \infty$, both bounds behave as $ 1/ \lambda $ while
for 
%AD: Changed symbol to \to
$\lambda \to 0$, the upper bound diverges while the lower bound
goes to 0 linearly in $\lambda$. The behavior of $\kappa_{GK}$ and of $\kappa$
in the NESS when $\lambda \rightarrow 0$ after $N \rightarrow \infty$ is thus
not determined by these bounds and remains an open problem for both the pinned
and unpinned random mass case.    
What we do know is that, if  $\lambda \rightarrow 0$ with N finite then there is a significant difference between the pinned and
unpinned cases \cite{dharleb08,dhar01}. As already noted in the introduction, for the pinned case all phonon modes are localized with a
fixed localization length independent of $N$ and
the current decays exponentially with system size. In the unpinned 
case the low frequency modes are extended and the current has a power
law decay with an exponent that depends on the boundary conditions
used \cite{dhar01}, $\kappa_N \sim N^{-1/2}$ for fixed BCs \cite{casher71,AH10} and as
$\kappa_N \sim N^{1/2}$ for free
BCs \cite{rubin71,verheggen79}. With the addition of the noisy dynamics which conserves
energy but not momentum we expect as noted earlier that the conductivity $\kappa$ will be equal to $\kappa_{GK}$  and thus strictly positive for any $\lambda >0$ \cite{BernOlla05}.   
 In the  following section we  evaluate $ \la J_{N} \ra$ as a function of $\lambda$ and N numerically and via computer simulations, to obtain information about its behavior when $\lambda \rightarrow 0$. 
 
\section{Results from numerics and simulations}
\label{numerics}

We study the dependence of the  heat current in the NESS on the system
 size and on the strengths of the   disorder and noise. 
In all our computations we set $k=1$. The masses
$\{m_l\}$ are chosen  from a uniform distribution between $1-\Delta$
to $1+\Delta$. This gives $[m]=1, [{1}/{m}]= 1/(2 \Delta) \ln [(1 +
 \Delta)/(1 - \Delta)]$. The average heat current 
%AD: Corrected mass index from l to l+1
 from site $l$ to $l+1$ is  given by $j_{l+1,l}= k \la q_l p_{l+1}/{m_{l+1}}
 \ra$. In the  steady state this is independent of $l$ and we denote
 $j_{l+1,l}=\la J_N \ra$. We note that $\la J_N \ra= k \hz_{l, l+1}/m_{l+1}$  and hence we
 can obtain accurate numerical values for the current in the
 disordered system by solving the  equations for the correlation matrix
 {\emph{i.e}} Eqs.~(\ref{nesseqs}). This involves solving large
 dimensional linear matrix  equations and we have been able to do this for
 system sizes less than $N=512$. For larger sizes we performed
 nonequilibrium simulations and obtained the steady state current by
a time average. For small sizes we verified that both methods agreed
 to very high accuracy. 
The number for disorder realizations  was  $100$ for $N \leq 64$, 
and varied between $2-16$ for larger sizes.  The error bars in our
 data presented below  are  calculated using the results from different realizations. 
  
\subsection{Pinned case}
\begin{figure}[b]
\vspace{1cm}
\includegraphics[width=5 in]{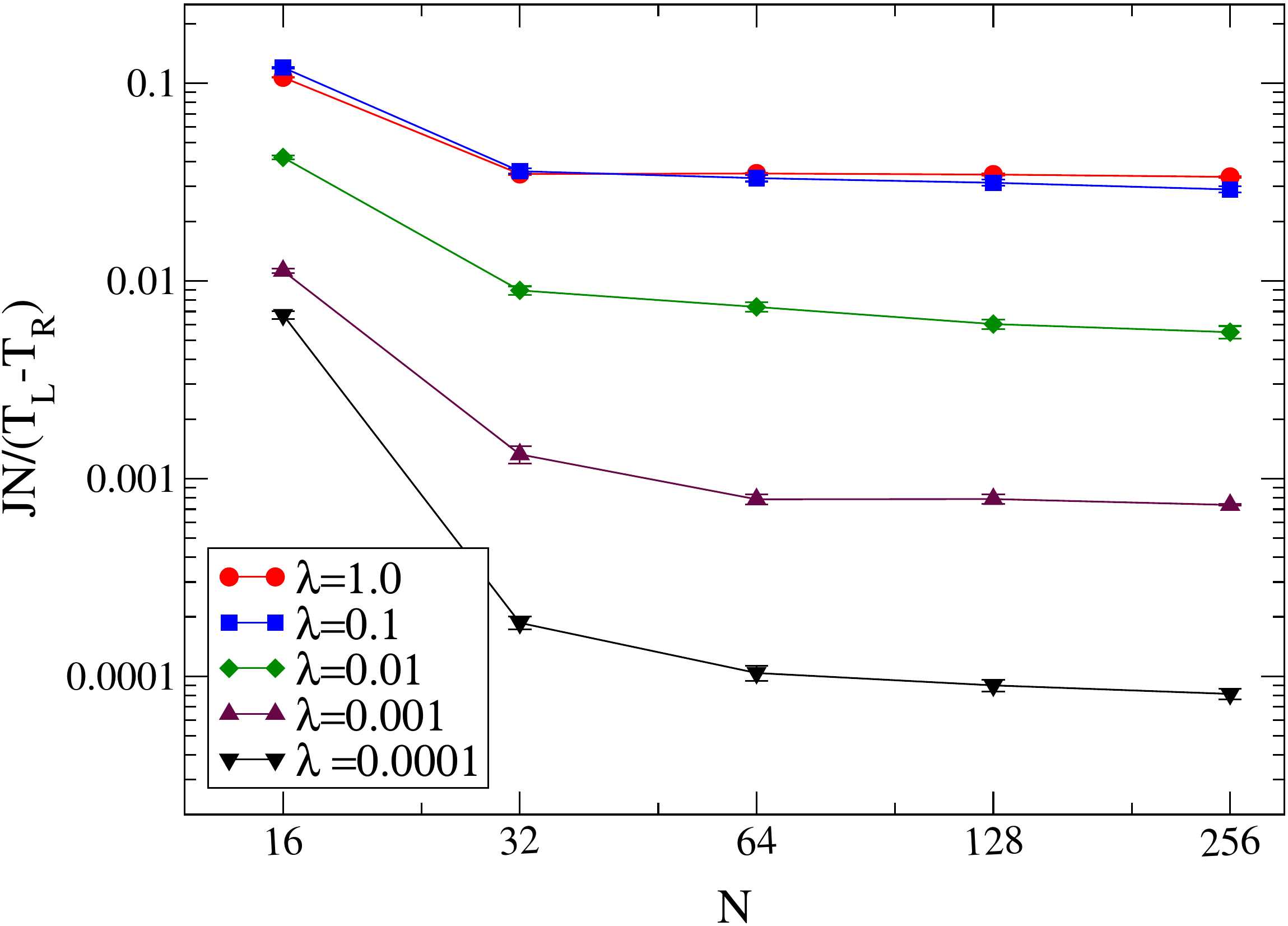}
\caption{(Color online)Plot of  $JN/(T_L-T_R)$ versus $N$ for for different values of
  $\lambda$. The parameter values were set  at $k_o=4, k=1,
  \Delta=0.8$.  All  the data shown here were obtained from exact numerical
  computation. }  
\label{pin-kap-vs-n}
\end{figure}

\begin{figure}
\vspace{1cm}
\includegraphics[width=4 in]{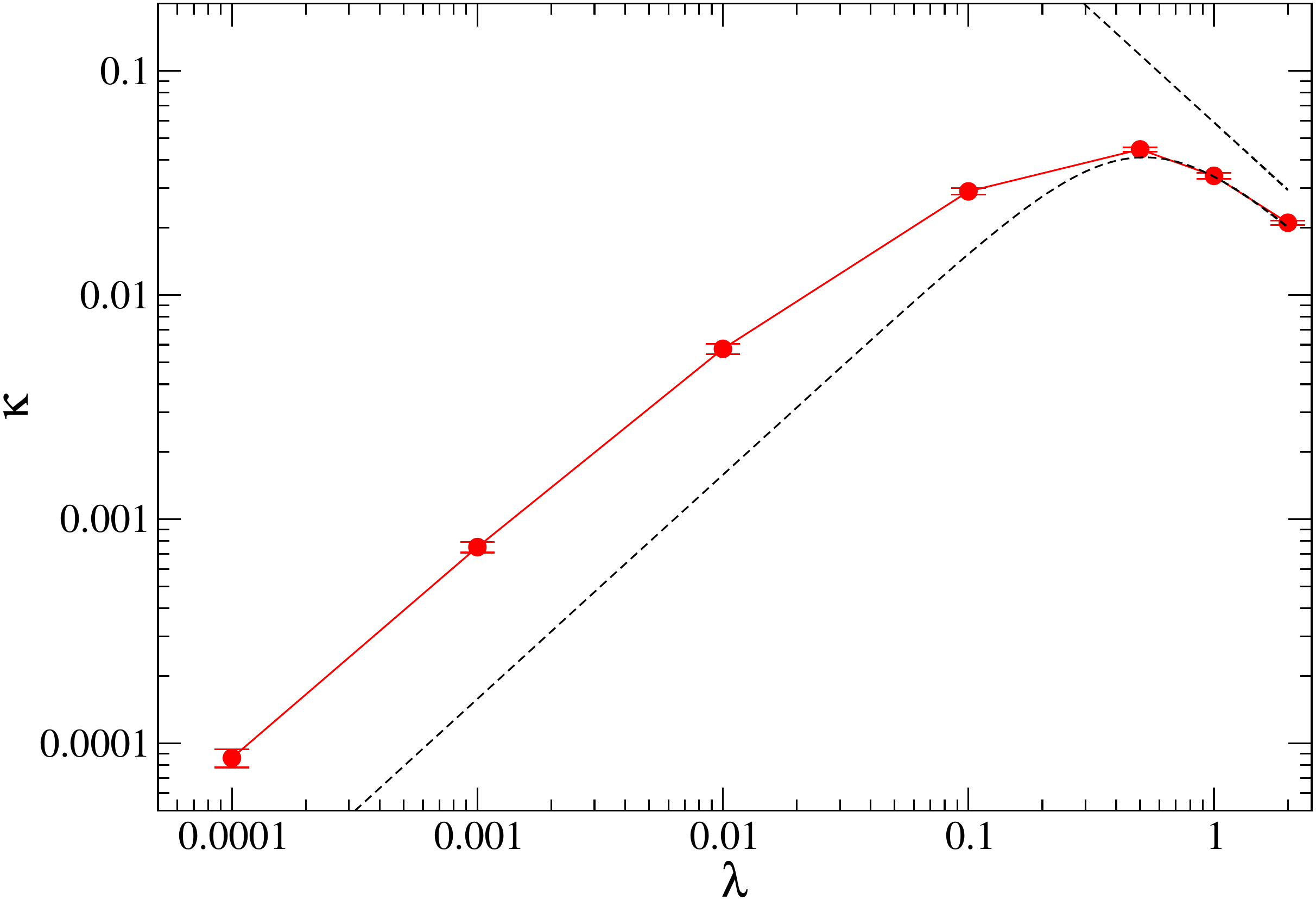}
\caption{(Color online) Plot of $\kappa$ versus $\lambda$ obtained from the
  numerical data in 
  Fig.~(\ref{pin-kap-vs-n}). The lower and upper bounds for
  $\kappa_{GK}$ given by   Eqs.~(\ref{eqlb},\ref{equb}) are shown by the dashed lines.}  
\label{pin-kap-vs-alp}
\end{figure}

This corresponds to the case with $k_o>0$ and here we also set $k'=1$. All
results in this section were obtained
by numerical solution of Eqs.~(\ref{nesseqs}).
In Fig.~(\ref{pin-kap-vs-n}) we plot $\la J_{N} \ra N/\Delta T$ versus
$N$ for different 
values of the flipping rate $\lambda$ and with $\Delta=0.8,~k_o=4$. In all
cases we see a rapid convergence 
to a system-size independent value which then gives the conductivity $\kappa$
of the system.  In Fig.~(\ref{pin-kap-vs-alp})
we plot $\kappa$, obtained from the large-$N$ data in  
Fig.~(\ref{pin-kap-vs-n}), as a function of $\lambda$. For comparison we also
plot the lower and upper bounds for the Green-Kubo conductivity given by
Eqs.~(\ref{eqlb},\ref{equb}). It is seen that $\kappa$ has a maximum around $\lambda \simeq 0.5$. This can be thought of as a balance between the flips delocalizing the phonons and acting as scatterers of phonons.  

\begin{figure}
\vspace{1cm}
\includegraphics[width=5.2in]{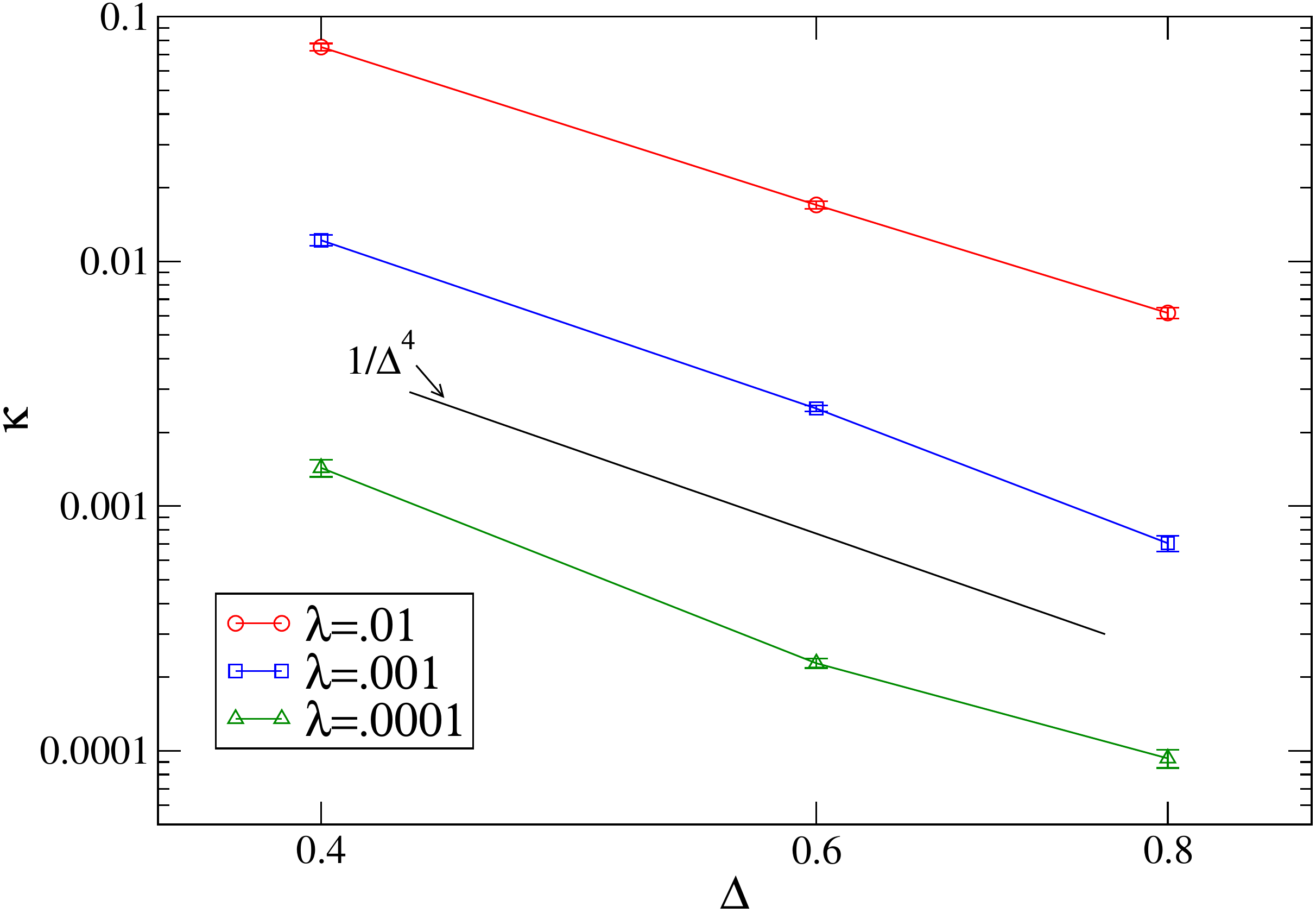}
\caption{ (Color online)Plot of $\kappa$ versus $\Delta$ for different values
  of $\lambda$ 
  and with $k_o=4$ and $k=1$. We
  also show a straight line with slope $-4.0$.}  
\label{pin-kap-vs-del}
\end{figure}

\begin{figure}
\vspace{1cm}
\includegraphics[width=4.8in]{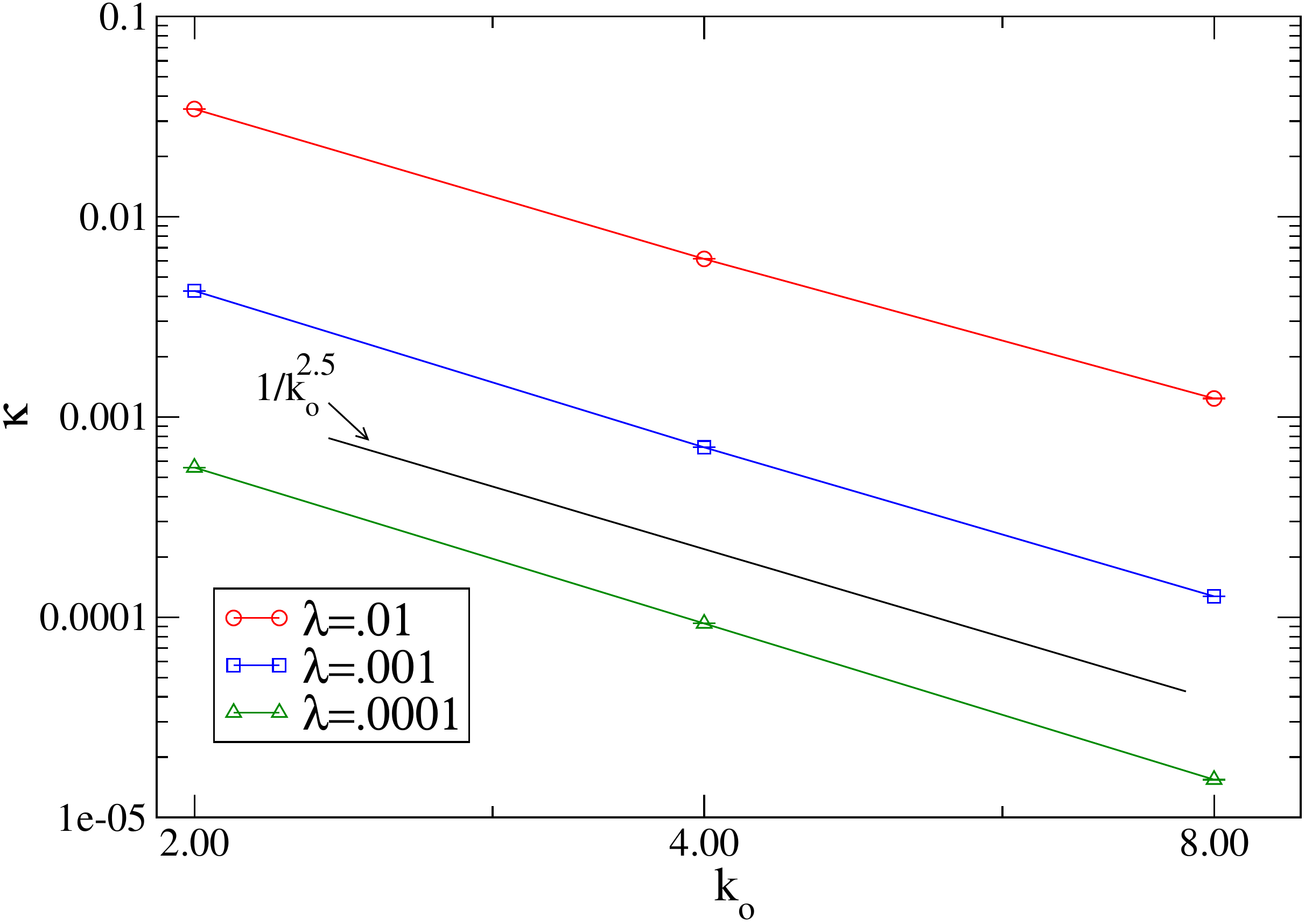}
\caption{(Color Online)Plot of $\kappa$ versus $k_o$ for different values of $\lambda$ and
  with $\Delta=0.8$ and $k=1$. We
  also show a straight line with slope $-2.5$.}  
\label{pin-kap-vs-K}
\end{figure}

In accord with the bounds we find that at  large $\lambda$, $\kappa \sim
1/\lambda$ while at small $\lambda$, the numerical results suggest $\kappa \sim \lambda$.
We note that for $\lambda=0$, all phonon modes are exponentially
localized within length-scales $\ell \sim (k_o \Delta^2)^{-1}$. One
can argue that for small values of $\lambda$ there is diffusion of
energy between these localized states with a diffusion constant $\sim
\ell^2 \lambda$. This leads to the  $\kappa \sim (k_o^2
\Delta^4)^{-1} \lambda$ and we now test this numerically.  
In Figs.~(\ref{pin-kap-vs-del},\ref{pin-kap-vs-K}) we show the numerical data
which suggests the scalings $\kappa \sim \Delta^{-4} $ and $\kappa
\sim {k_o}^{-2.5}$, which are roughly consistent with the expected behavior. The reason for the discrepancy could be that we
are not yet in the strong localization regime where the prediction is
expected to be most accurate.

\begin{figure}
\includegraphics[width=5.2in]{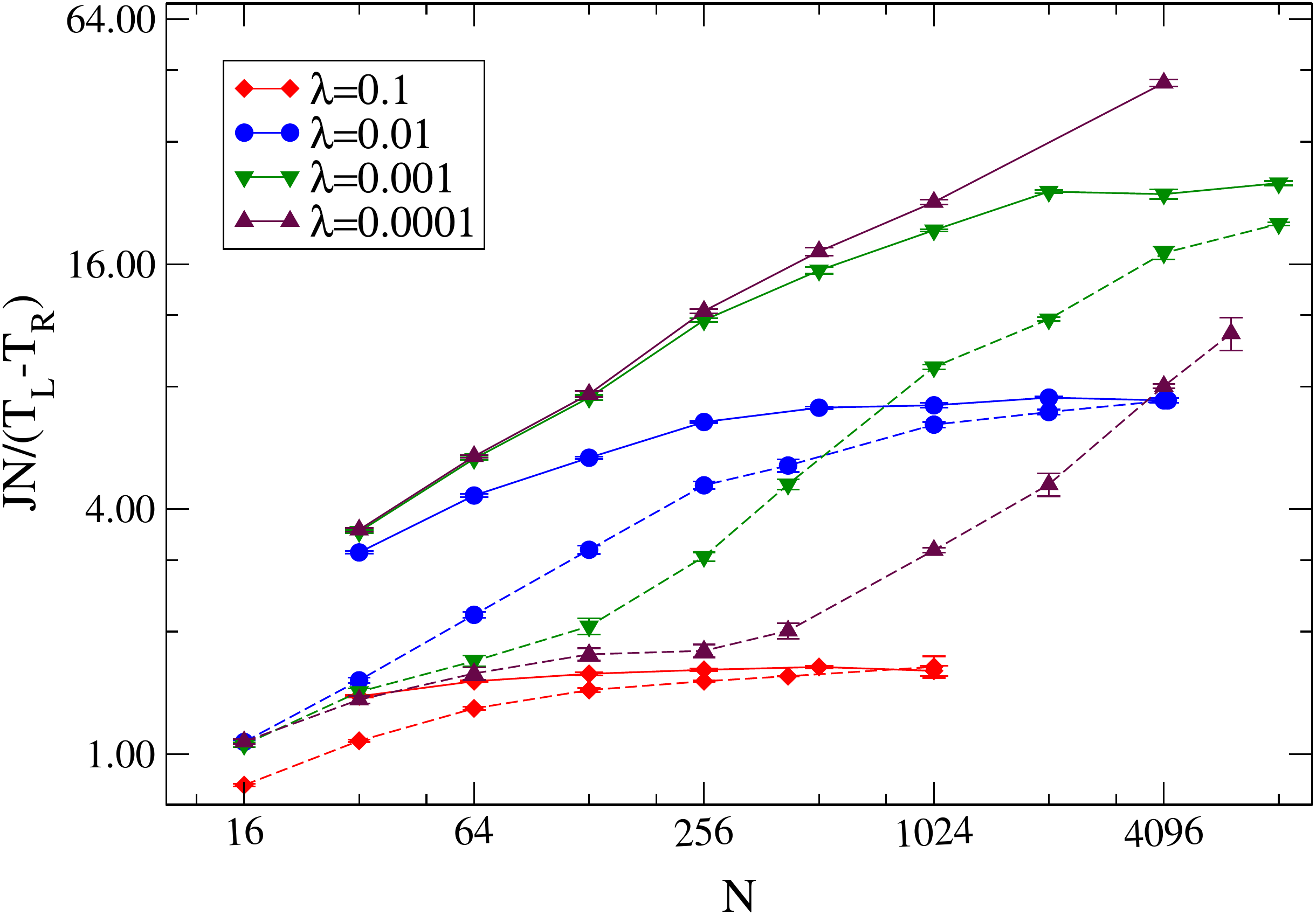}
\caption{ (Color Online)Plot of $JN/(T_L-T_R)$ versus $N$ for the unpinned case with
  both fixed (dashed lines) and   free BCs (solid lines) for   different values of $\lambda$ and parameter values $k=1$ and
  $\Delta=0.8$. The data for $ N < 512$ were obtained using exact numerics
  and in all these cases simulations give very good agreement with the
  numerics. For  $ N \geq 512$, the data 
  were obtained from   simulations alone. }    
\label{unpin-kap-vs-n}
\end{figure} 

\begin{figure}
\vspace{1cm}
\includegraphics[width=5.2in]{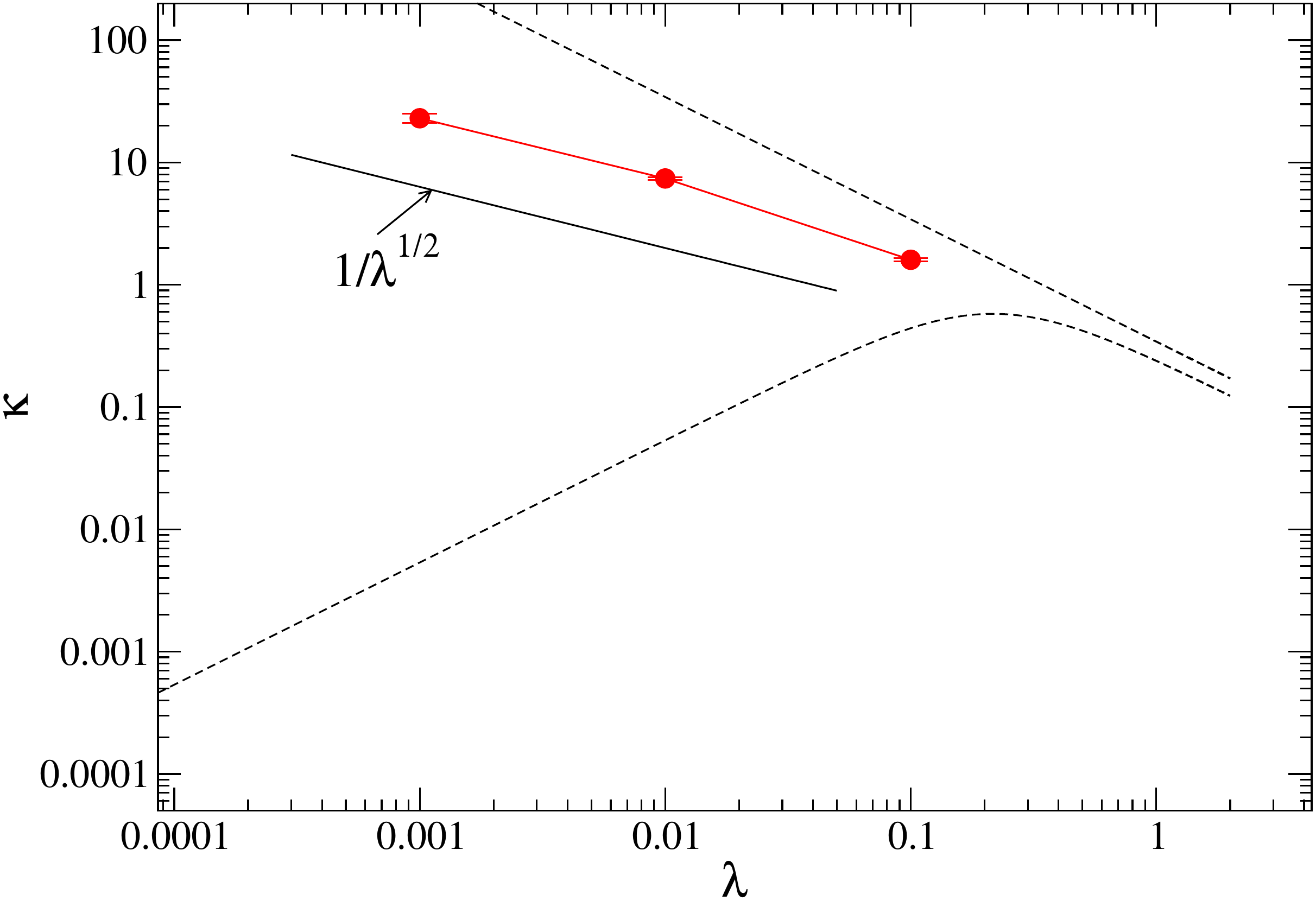}
\caption{(Color Online)Plot of $\kappa$ versus $\lambda$ for the unpinned system
  obtained from the numerical data in 
  Fig.~(\ref{unpin-kap-vs-n}). The lower and upper bounds for
  $\kappa_{GK}$ given by
  Eqs.~(\ref{eqlb},\ref{equb}) are shown by the dashed lines. Also
  shown is a straight line with slope $-1/2$.}  
\label{unpin-kap-lam}
\end{figure} 

 % \vspace{1cm}
\begin{figure}
\includegraphics[width=4.5 in]{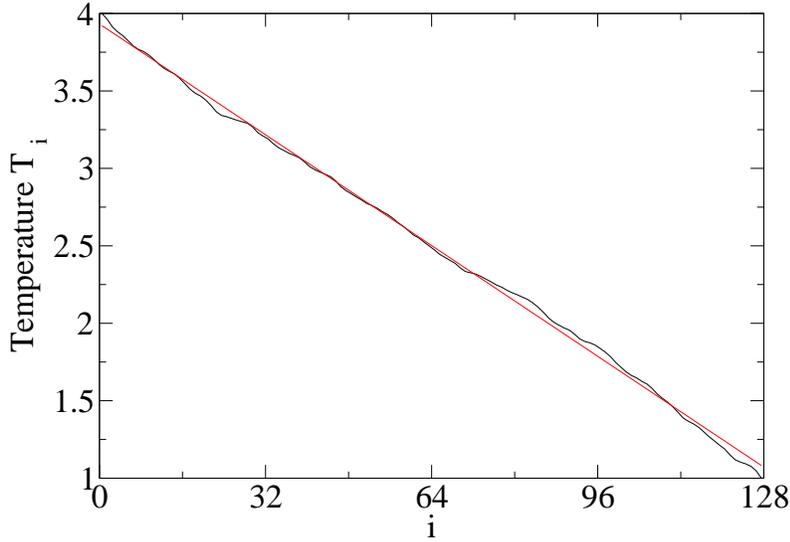}
 \caption{(Color Online)Plot of temperature profile ($T_i=\la p_i^2/m_i \ra$) for the
  unpinned case (with mass disorder $\Delta =0.8$) with fixed BCs for 
  $N=128, \lambda=5, T_{L}=4, T_{R}=1$. The expected linear
  profile is also shown.  The data was obtained from exact numerical
  computation.} 
\label{Temp-prof128}
\end{figure}

\begin{figure}
% \vspace{1cm}
\includegraphics[width=4.5 in]{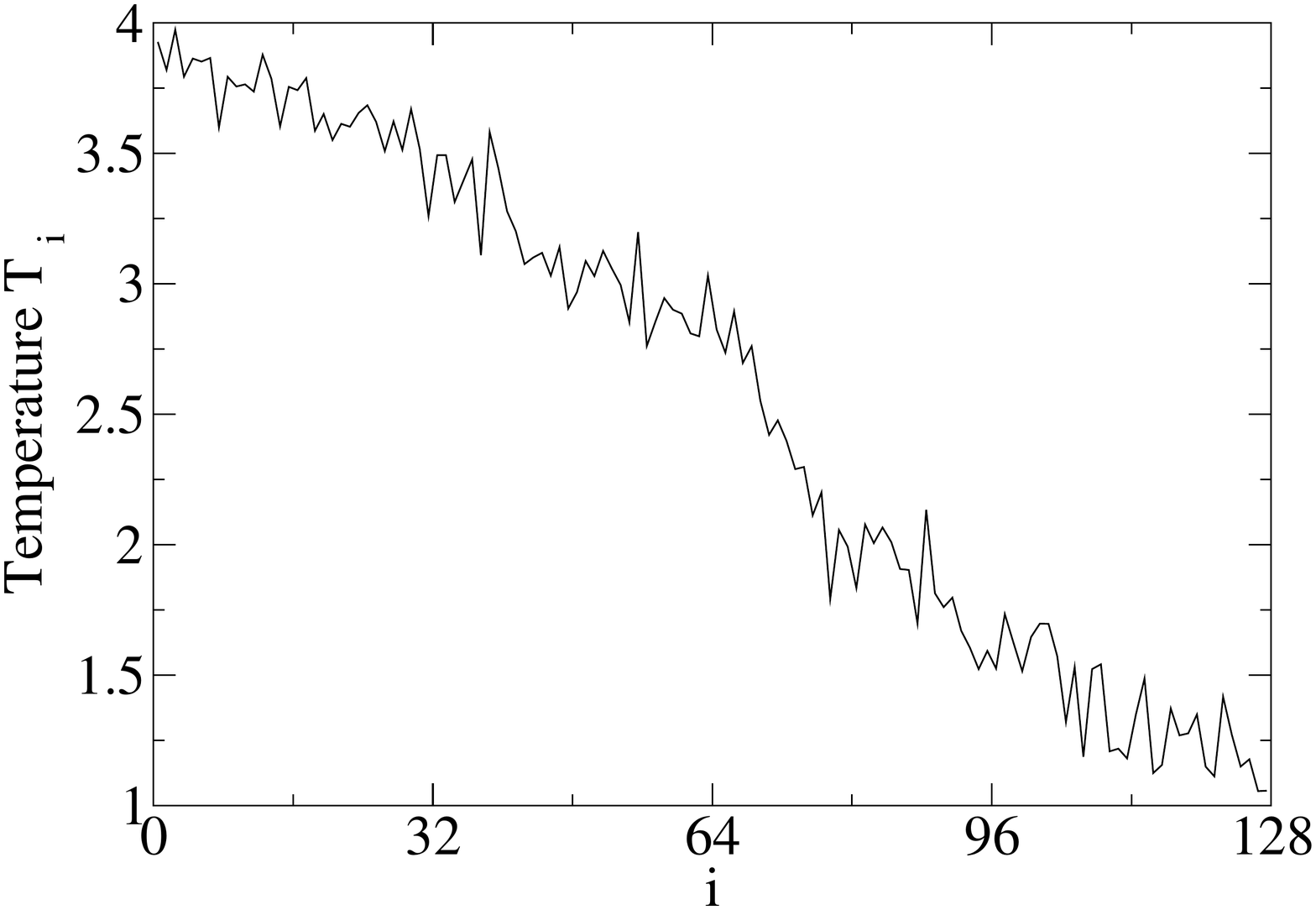}
 \caption{ Same parameters as Fig. \ref{Temp-prof128} except the velocity flip
   $\lambda \rightarrow 0$. The data shown is from an exact numerical computation with
$\lambda=10^{-9}$. We have verified that  this is close to the temperature
profile for $\lambda=10^{-7}$ and expect that it is converging to the
$\lambda=0$ value.}
\label{Temp-prof128D}
\end{figure}
\subsection{Unpinned case}
As noted above for the unpinned case with $\lambda=0$,  the two
different boundary conditions(BC) namely fixed BCs with $k' > 0$ and free
BCs with $k'=0$ give respectively $\la J_{N} \ra \sim N^{-3/2}$
\cite{casher71} and $ \la J_{N} \ra \sim N^{-1/2}$ \cite{rubin71}. The difference
in the asymptotic behavior of the current for 
different BCs can be understood as arising from the dependence on BCs
of the transmission of the low frequency modes which carry the current \cite{dhar01}.  
For any $\lambda > 0$ however we expect that the system should have a
unique finite value of the conductivity, independent of boundary
conditions, the same as for $\kappa_{GK}$. Physically  we can argue
as follows: 
The unpinned system without disorder has a finite
positive $\kappa$ given by Eq.~(\ref{kGK}), which is independent of BCs [see
comments at end of sec.~(\ref{subsec:eqmass})]. The low frequency modes 
are weakly affected by disorder hence we expect that as far as these modes are
concerned the unpinned system with and without disorder will behave
similarly. Since these are the modes which led to the dependence on BCs for
the case $\lambda =0$, we expect that for $\lambda > 0$ they will not have any
effect.  

\begin{comment}
\begin{figure}
% \vspace{1cm}
\includegraphics[width=3 in,angle=-90]{pro1024s05.pdf}
\caption{ $N=1024, \lambda=0.5, T_{L}=4, T_{R}=1$. }
\label{Temp-prof}
\end{figure}
\end{comment} 

We now present results of our numerical and simulational studies of the
unpinned chain with free and fixed BCs. The numerical results are obtained
by solving Eqs.~(\ref{nesseqs}). The simulation involves evolving the
system  with the Hamiltonian part, the
 momentum flips at all sites and the
Langevin baths at the boundary sites. For $N \leq 512$, the numerical
method was employed to arrive at the solution for the NESS whereas for
larger values of $N$, we performed simulations to obtain the data.   In
Fig.~(\ref{unpin-kap-vs-n}) we 
plot $JN/(T_L-T_R)$ versus $N$ for different values of $\lambda$ for both fixed and free BCs. 
The disorder strength is $\Delta=0.8$.  For both BCs we can see
flattening of the curves at large system sizes for   the parameter values
$\lambda =0.1, 0.01$ implying a finite $\kappa$, which is independent
of BCs. For $\lambda =0.001,0.0001$ it
appears that reaching the asymptotic limit requires larger system sizes.  
Using the large-$N$  data in Fig.~(\ref{unpin-kap-vs-n}) we
estimate the conductivity 
$\kappa =JN/\Delta T$ and this is plotted in Fig.~(\ref{unpin-kap-lam}).
In Fig.~(\ref{Temp-prof128}) we show a typical plot of the temperature
profile for the case with 
fixed BC (this was obtained using exact numerics). The profile is close to
linear consistent with the fact that the 
conductivity is temperature-independent. We do not see any significant
boundary temperature 
jumps since the system size is sufficiently large. In Fig  
(\ref{Temp-prof128D}) we have the profile for N=128 for the case  
$\lambda \to  0$ which shows considerable jump in the temperatures across
neighboring sites.        
\begin{table}
\caption{Values of Correlation Functions for n=4 ($T_L=4, T_R=1$)}
\begin{tabular} {|l|c|c|c|} 
\hline
Correlation & $\lambda =0.1 $ & $\lambda=2$ & $\lambda = 10 $  \\
\hline
$\f{<p_1^4> -3<p_1^2>^2}{<p_1^2>^2}$ & 0.006 & 0.014 & $ \sim 10^{-4}$ \\
$\f{<p_2^4> -3<p_2^2>^2}{<p_2^2>^2}$ & 0.054 & 0.058 & 0.049 \\
$\f{<p_3^4> -3<p_3^2>^2}{<p_3^2>^2}$ & 0.061 & 0.110 & 0.098 \\
$\f{<p_4^4> -3<p_4^2>^2}{<p_4^2>^2}$ & 0.038 & 0.018 & 0.002 \\
% $\f{<q_1^4> -3<q_1^2>^2}{<q_1^2>^2}$ & 0.001 & 0.018 & 0.004  \\ 
% $\f{<q_2^4> -3<q_2^2>^2}{<q_2^2>^2}$ & 0.011 & 0.059 & 0.079 \\
% $\f{<q_3^4> -3<q_3^2>^2}{<q_3^2>^2}$ & 0.015 & 0.086 & 0.193 \\
% $\f{<q_4^4> -3<q_4^2>^2}{<q_4^2>^2}$ & 0.002 & 0.097 & 0.011 \\

$\f{<p_1^2 q_1^2> -<q_1^2><p_1^2>}{<q_1^2><p_1^2>}$ & 0.002 & 0.004 & $\sim 10^{-5}$ \\
$\f{<p_2^2 q_2^2> -<q_2^2><p_2^2>}{<q_2^2><p_2^2>}$ & 0.011 & 0.014 & 0.013 \\
$\f{<p_3^2 q_3^2> -<q_3^2><p_3^2>}{<q_3^2><p_3^2>}$ & 0.012 & 0.025 & 0.023  \\
$\f{<p_4^2 q_4^2> -<q_4^2><p_4^2>}{<q_4^2><p_4^2>}$ & 0.012 & 0.034 & $ \sim 10^{-4}$ \\
% $\f{<p_1^2 p_3^2> -<p_1^2><p_3^2> -2<p_1 p_3>^2}{<p_1^2><p_3^2>}$ &  0.000 & -0.002 & 0.000 \\ 
% $ \f{<p_3^2 p_4^2> -<p_3^2><p_4^2> -2<p_3 p_4>^2}{<p_3^2><p_4^2>}$ & 0.000 & 0.138 & 0.111  \\
\hline
\end{tabular}
\end{table}
It appears likely that for all $\lambda > 0$ the
conductivity $\kappa$ is independent of BCs.  However this is difficult
to verify from simulations since one needs to study very large system
sizes to reach the correct asymptotic limit. 
The reason for this can be roughly seen as follows. In the ordered case the
conductivity 
$ \kappa \sim 1/\lambda$ and this can be understood in terms of an
effective mean free path $\ell \sim 1/\lambda$ for the ballistic
phonons because of scattering from the stochastic process.  
Hence we can expect that, to see diffusive behavior for the low
frequency ballistic modes,  important in the disordered
case,  requires one to study sizes $N \stackrel{>}{\sim} \ell$ or $N
\stackrel{>}{\sim}1/\lambda$. 
Finally we observe from Fig.~(\ref{unpin-kap-lam}) that at small $\lambda$, the
conductivity appears to be diverging as $1/\lambda^{1/2}$. 
In the absence of noise the localization length $\ell_L \sim
1/\om^2$, hence  it is expected that all modes with $\ell_L < \ell$
or $\om > \lambda^{1/2}$ stay localized.  The low frequency modes
$0< \omega < \lambda^{1/2}$ become  diffusive with mean free paths
$\sim 1/\lambda$ thus resulting in a conductivity $\kappa \sim
\lambda^{1/2} (1/\lambda) \sim 1/\lambda^{1/2}$, which explains the
observed behavior.

\section{Discussion}
\label{summary}

We have shown here  that the stationary one-body and pair correlations
in the velocity flip model are the same (after setting $\gamma'_{l}=2
\lambda m_{l}$) as in the harmonic chain with 
self-consistent reservoirs. 
We have also numerically investigated the dependence of the thermal
conductivity $\kappa$ of the
disordered harmonic chain on the velocity flip rate $\lambda$. For $\lambda \to
0$ our results suggest $\kappa \sim \lambda$ for the pinned system
and $\kappa \sim \lambda^{-1/2}$ for the unpinned system. Establishing
these results conclusively requires further work. 

We note  that while for the self-consistent reservoirs model  
the  NESS is exactly Gaussian, this is not so for our
noisy model. Instead it will be in general a superposition of
  Gaussians. Computer simulations however indicate that the
 single particle distributions are very close to a single Gaussian
 while the joint distribution of $x_{l}$ and $p_{l}$ or of $p_{l}$ and
 $p_{j}$ , $ j \neq l$, are essentially uncorrelated. In addition, the set of equations for the four variable correlation functions was derived and for small values of n, they were solved numerically. The exact values obtained from these numerics were found to be in close match with the simulation results. For n=4, and $T_L=4, T_R=1$, the results from the numerical solution are shown in table I corresponding to 3 different values of $\lambda$. It was also observed that these normalized forms of correlations (that go to zero when the temperatures of the two reservoirs are equal or when there is no velocity flipping) have a limit when either the difference in temperature ($T_{L}-T_{R}$) or strength of stochastic noise ($\lambda$) goes to infinity. When we let $\lambda \to \infty$, we observed that the correlations involving p1 and p4 went to zero. \\ \\
We are currently investigating the $O(N)$ corrections to the pair
correlations in the noisy NESS . These are known  to behave like $1/N$
for certain diffusive lattice systems and to contribute terms of $O(N)$
beyond those obtained from the local equilibrium to the variance of
the particle number in the NESS.  Results of this kind are also known
partially for the continuum case with a  different kind of noise, i.e
instead of velocity reversals pairs of nearest neighbor particles
diffuse on the circle $p_{i}^{2} + p_{i+1}^{2}=C$.

\section{Acknowledgements}
We thank Cedric Bernardin, David Huse, Jani Lukkarinen and Stefano Olla for very helpful discussions. Joel Lebowitz and Venkateshan K were supported by NSF grant DMR 08-02120 and by AFOSR[grant AF-FA09550-07]. Joel Lebowitz also thanks the IHES and IHP for their hospitality during part of this work.

\end{document}